\documentclass[superscriptaddress,aps,pra,twocolumn,showpacs,preprintnumbers,amsmath,amssymb,floatfix,groupedaddress]{revtex4-1}
\usepackage{graphicx}% Include figure files
\usepackage{dcolumn}% Align table columns on decimal point
\usepackage{bm}% bold math
\usepackage{color}
\usepackage{psfrag}
\begin{document}
\def\e{\mathcal{E}}
\def\ds{\delta_s}
\def\Dhf{\Delta_{\rm hf}}
\def\DR{\Delta_{\rm R}}
\def\Ge{\Gamma_{\rm E}}
\newcommand{\ket}[1]{|{#1}\rangle}
\newcommand{\bra}[1]{\langle {#1}|}
%\renewcommand{\exp}[1]{{\rm e}^{#1}}

%\title{Light storage in a $\Lambda$-system under conditions of electromagnetically induced transparency and four-wave mixing}
\title{Light storage in an optically thick atomic ensemble under conditions of electromagnetically induced transparency and four-wave mixing}

\author{Nathaniel B. Phillips$^{1}$}
\author{Alexey V. Gorshkov$^{2}$}
\author{Irina Novikova$^{1}$}
\affiliation{$^{1}$Department of Physics, College of William and Mary, Williamsburg,Virginia 23185, USA,\\
$^{2}$Institute for Quantum Information, California Institute of Technology, Pasadena, California 91125, USA}
%
%\affiliation{Department of Physics, College of William and Mary, Williamsburg,Virginia 23185, USA}

\date{\today}

\begin{abstract}
%We study the modification of a traditional electromagnetically induced transparency (EIT) stored light technique under EIT and four-wave mixing (FWM) conditions in an ensemble of hot Rb atoms.
We study the modification of a traditional electromagnetically induced transparency (EIT) stored light technique that includes both EIT and four-wave mixing (FWM) in an ensemble of hot Rb atoms. 
The standard treatment of light storage involves the coherent and reversible mapping of one photonic mode onto a collective spin coherence.  It has been shown that unwanted, competing processes such as four-wave mixing are enhanced by EIT and can %efficiently and spontaneously produce a new optical mode.
significantly modify the signal optical pulse propagation.  We present theoretical and experimental evidence to indicate that while a Stokes field is indeed detected upon retrieval of the signal field, any information originally encoded in a seeded Stokes field is not independently preserved during the storage process.  We present a simple model that describes the propagation dynamics of the fields and the impact of FWM on the spin wave.
\end{abstract}

\pacs{42.50.Gy, 32.70.Jz, 42.50.Md}

\maketitle
%%%%%%%%%%%%%%%%%%%%%%%%%%%%%%%%%%%%
%%%%%%%%%%%%%%%%%%%%%%%%%%%%%%%%%%%%
\section{Introduction}
%%%%%%%%%%%%%%%%%%%%%%%%%%%%%%%%%%%%
%%%%%%%%%%%%%%%%%%%%%%%%%%%%%%%%%%%%

The successful development of practical quantum information applications relies in large extent on the availability of high-efficiency and high-fidelity memory for quantum states of photons.  Recently, several promising realizations of such a quantum memory were demonstrated, which are based on the reversible mapping of photon quantum states onto long-lived collective coherences in ensembles of atoms~\cite{lukin03rmp,kimbleNature08,lvovskyNaturePh09,polzikRMP10,euromemory}.
In the majority of these protocols, an atomic ensemble of sufficiently high optical depth is necessary in order to achieve high memory efficiency. However, in this case, some unwanted %nonlinear 
competing processes may interfere with quantum memory performance, reducing its efficiency and fidelity \cite{lukinPRL97, kangFWM, narducciFWM, haradaFWM, agarwalFWM, taoFWM, phillipsJMO09}.

\begin{figure}[t]
	\center{\includegraphics[width=\columnwidth]{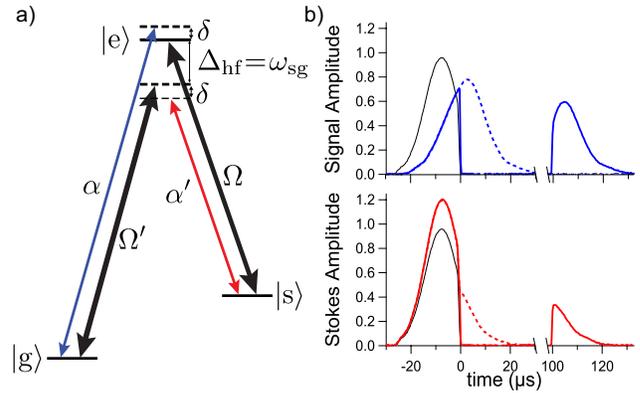}}
	\caption{\label{fig:doubleLambda} (Color online) (a) The double-$\Lambda$ system used in theoretical calculations.  In our case, $\ket{g}$ and $\ket{s}$ correspond to the $^{87}$Rb ground state hyperfine sublevels $|F,~m_F\rangle=|1,1\rangle,~|2,1\rangle$, respectively; $\ket{e}$ corresponds to the excited state $|F', ~m_{F'}\rangle=|2,2\rangle$.  $\Omega$, $\Omega'$ represent the Rabi frequencies of the same control field acting on two different transitions, while $\alpha$ and $\alpha'$ represent the Rabi frequencies of the signal and Stokes fields, respectively.  (b) %and (c) 
	Sample data for light storage for signal (top) and Stokes (bottom) pulses %correspondingly 
	at a temperature $T=70^{\circ}$C (optical depth $\alpha_0 L=52$). Dashed lines show propagation of these pulses under slow light regime (constant control field). The black curve is a far-detuned reference pulse.}  %\textit{Top} Signal.  \textit{Bottom} Stokes.  }
%graph20_2 in workd0720.pxp; T=70, \delta=20kHz (cancels stark shift)  \Omega=6.2 MHz
\end{figure}

In this manuscript, we investigate the propagation and storage of weak optical signal pulses in optically thick hot atomic vapor based on electromagnetically induced transparency (EIT)~\cite{lukin03rmp,fleischhauerRMP05}. Traditionally, an EIT-based light storage scheme considers the simultaneous interaction of a strong control  field and a weak signal %optical 
fields in a $\Lambda$-type configuration, in which two ground hyperfine levels are linked with a common excited state, as shown in Fig.~\ref{fig:doubleLambda}(a). In this case, the control field strongly couples the propagation of the signal optical field with a collective long-lived ground state atomic spin coherence (spin wave) \cite{fleischhauer,lukin03rmp}, resulting in a reduced group velocity for signal pulses (``slow light''). Adiabatic turn-off of the control field maps the quantum state of the signal field onto the spin wave, which can be stored and later retrieved by restoring the control field intensity.

Such EIT quantum memory has been realized for both weak classical and non-classical signal pulses (for recent reviews, see Refs.\ \cite{lvovskyNaturePh09,polzikRMP10,euromemory}). Some recent publications~\cite{gorshkovPRL,novikovaPRLopt,phillipsPRA08} investigated the optimal performance of such an EIT memory and have confirmed that high optical depth is necessary to improved storage efficiency. For example, $90\%$ memory efficiency requires an optical depth $\alpha_0 L > 100$~\cite{gorshkovPRA2}. On the other hand, an optically dense coherent atomic medium is also known to enhance competing undesired %nonlinear 
effects, such as resonant four-wave mixing (FWM) 
in a double-$\Lambda$ configuration. In this FWM process, the far-detuned interaction of the control field, which resonantly drives
%tuned in resonance with 
the $|s\rangle-|e\rangle$ transition [see Fig.~\ref{fig:doubleLambda}(a)], 
with atomic ground state coherence enhances %nonlinear 
the generation of an off-resonant Stokes optical field [$\alpha'$ in Fig.\ \ref{fig:doubleLambda}(a)]. In turn, the presence of a Stokes field strongly affects signal pulse propagation, and the propagation of both signal and Stokes fields become a result of the interference between regular EIT and FWM processes~\cite{phillipsJMO09, eugeniyFWMJMO, taoFWM}.

Under such conditions, the simplified treatment of an EIT-based quantum memory in a single-$\Lambda$ system is incomplete and fails to describe light storage at high optical depths~\cite{phillipsPRA08}. In this work, we explore the mechanisms governing propagation of signal and Stokes pulses under EIT-FWM conditions and develop an intuitive analytical treatment of this problem. In particular, we investigate the prospect of their light storage, %\textit{e.g.}, 
\textit{i.e.}, a process in which both signal and Stokes pulses are reversibly mapped onto a long-lived spin coherence and thus can be faithfully recreated after some storage period.  Recent experiments~\cite{howellNatPh09} have showed that a spontaneously-generated Stokes field can be detected upon retrieval of a signal field from a spin coherence.  Based on these results, one might anticipate that the spin wave might function, at least to some extent, as a memory for both pulses. In our experiments, both input signal and Stokes fields have non-zero amplitudes before entering the atomic medium. We observed that under certain experimental conditions, both signal and Stokes pulses appear to be delayed due to the interaction with atoms, \textit{i.e.}, both signal and Stokes outputs are nonzero even after the input signal and Stokes fields are turned off. Moreover, when the control field is turned off for some time and then later turned on, output pulses are retrieved in both channels, as shown in Fig.~\ref{fig:doubleLambda}(b). However, careful experimental and theoretical investigation shows that this is not a two-mode storage, \textit{i.e.}, the signal and Stokes fields  cannot be stored independently.  Instead, under EIT-FWM conditions, the collective ground-state spin coherence is determined by a particular combination of signal and Stokes fields. Moreover, both retrieved fields are, in fact, only very weakly sensitive to the input Stokes field.  %Furthermore, 
To explain these effects, we present an intuitive analytical picture of the effects of FWM on the signal and Stokes pulses and on the atomic spin coherence.
%
%We found that for low and moderate optical depth of atomic ensemble ($\alpha_0 L \leq 25$) it is still possible to describe the coupled propagation of signal and Stokes pulses using the dark-state polariton formalism~\cite{fleischhauer} by replacing a signal field in the traditional single-$\Lambda$ EIT by this joint mode combination of two fields. Under these conditions, it seems possible to store and retrieve this combination of signal and Stokes fields with good fidelity, even though each individual field is not independently preserved. However, for higher optical depth, this treatment fails due to additional gain or absorption of the Stokes field.
%
%\textbf{not sure what to say next.}  The effect of four-wave mixing can be advantageous or detrimental, depending on the details of the application. On one hand, in quantum memory applications, the resonant mixing reduces the fidelity by adding extra noise into the signal field. On the other hand, non-classical correlations between two signal and Stokes fields can individually carry quantum information ~\cite{lett} and produce entangled images. Also, under certain conditions, FWM may lead to gain for both the signal and Stokes fields, which could compensate for any optical losses during slow and stored light processes~\cite{howellNatPh09}.

%%%%%%%%%%%%%%%%%%%%%%%%%%%%%%%%%%%%
%%%%%%%%%%%%%%%%%%%%%%%%%%%%%%%%%%%%
\section{Experimental Arrangements}
%%%%%%%%%%%%%%%%%%%%%%%%%%%%%%%%%%%%
%%%%%%%%%%%%%%%%%%%%%%%%%%%%%%%%%%%%

\begin{figure}[bt]
	\center{\includegraphics[width=0.5\textwidth]{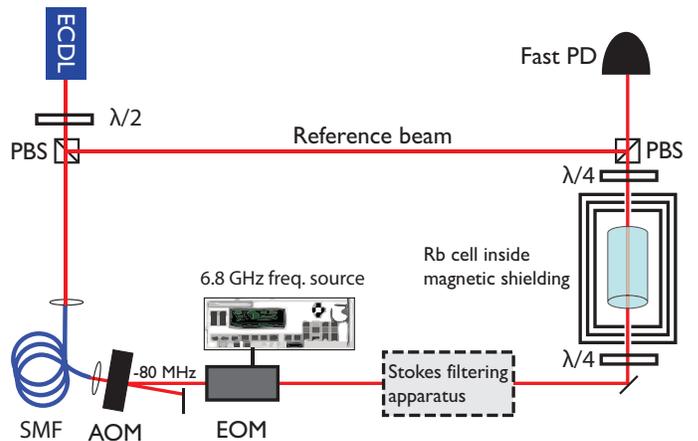}}
	\caption{\label{fig:setup} (Color online) A schematic of the experimental arrangements.  See text for abbreviations.}
\end{figure}

A schematic of the experimental setup is shown in Fig.\ \ref{fig:setup}.  The output from an external cavity diode laser (ECDL) locked to the Rubidium $D_1$ transition ($\lambda=795$ nm) passed through a polarizing beam splitter (PBS), which split off a fraction of the beam for use as a reference frequency, while the rest of the beam was coupled into a single-mode optical fiber (SMF) to improve its transverse intensity distribution. All experimental data shown below were obtained using weak classical laser pulses. To ensure the best spatial overlap and mutual phase coherence, all three optical fields were derived from a single laser beam by phase modulation. After the fiber, the beam passed an acousto-optical modulator (AOM), and the $-1$ diffraction beam (downshifted by $80$ MHz) then passed through an electro-optical phase modulator (EOM) operating at the frequency of the ground state hyperfine splitting of $^{87}$Rb [$\Dhf/(2\pi)=6.835~$GHz].  This phase modulation produced two first modulation sidebands at $\pm\Dhf$ of nearly equal amplitudes and opposite phases.  The zeroth order (carrier frequency) field was tuned to the $5^2$S$_{1/2}$F$=2\to5^2$P$_{1/2}$F$'=2$ %used to say S
 transition and acted as the control field $\Omega$~\cite{Omeganote}.  The $+1$ modulation sideband, tuned near the $5^2$S$_{1/2}$F$=1\to5^2$P$_{1/2}$F$'=2$, %used to say S
 played the role of the signal field, while the $-1$ sideband acted as the % far-detuned
Stokes field.  The resulting beam was %were 
collimated to a diameter of $1.9$ mm or $2.7$ mm, as specified below, and circularly polarized with a quarter-wave plate ($\lambda/4$).  Typical peak control and signal field powers were approximately $14$ mW and $40~\mu$W, respectively.  For some experiments, we attenuated the amplitude of the Stokes field as described in Refs.\ \cite{phillipsJMO09, phillipsPRA08}.

A cylindrical Pyrex cell, of length $75$ mm and diameter $22$ mm, containing isotropically enriched $^{87}$Rb and 30 Torr Ne buffer gas, so that the pressure-broadened optical transition linewidth was $2\gamma/(2\pi)= 290$ MHz \cite{Rotondaro}, was mounted inside tri-layer $\mu$-metal magnetic shielding, in order to reduce the effects of stray magnetic fields.  The temperature of the cell, and correspondingly the concentration of Rb in the vapor phase, was controlled using a bifilar resistive heater wound around the inner-most shield layer.  Experimental temperatures ranged from $50^\circ$C to $80^\circ$C, which corresponded to changes in Rb density from $1.1\times10^{11}$ cm$^{-3}$ to $1.2\times10^{12}$ cm$^{-3}$, and to a range of optical depths $\alpha_0 L$ between 10 and 110.
After the cell, the light beam was %were 
linearly polarized with a $\lambda/4$ plate, recombined with the unshifted reference beam, and sent via a multi-mode optical fiber to a fast photodetector (PD).  The beat note signals between each of the $+1$ and $-1$ modulation sidebands and the reference field were measured using a microwave spectrum analyzer.  Because of the $80$ MHz frequency shift introduced by the AOM, the different beat note frequencies of each sideband with the reference field allowed for independent measurement of their amplitudes.

Simultaneously-programmed modulation strengths of the AOM and EOM allowed for independent control over the temporal envelopes of the control field and the signal/Stokes fields. In this set of measurements, we used constant levels of the control field for both writing and retrieval stages and truncated Gaussian waveforms for the signal and Stokes fields. Each slow and stored light measurement was preceeded by a $400~\mu$s pulse of the control field at maximum intensity to ensure optical pumping of the atoms in the interaction region into state $\ket{g}$.

\begin{figure}[tb]
\center{\includegraphics[width=0.9\columnwidth]{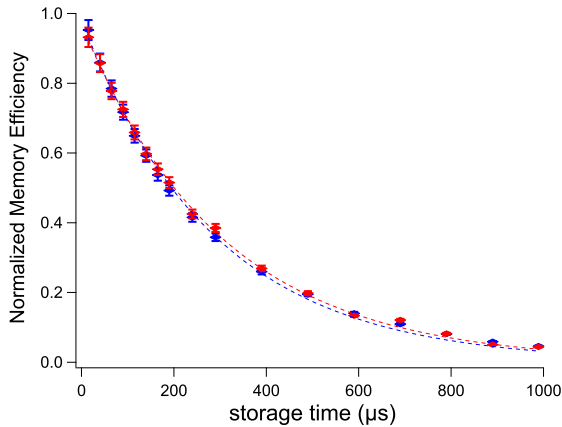}}
\caption{\label{decaygraph}(Color online) Dependence of retrieved signal (blue) and Stokes (red) pulse energies as a function of storage time at $T=70^\circ$C ($\alpha_0 L=52$).  Here, $\tau_s=300~\mu$s.  We normalized the memory efficiencies so that the zero-storage-time memory efficiency is unity.  Experimentally, we were not operating under optimal storage conditions \cite{phillipsPRA08}, and the zero-storage-time memory efficiency was $\approx 40\%$ for the signal pulse and $\approx 5\%$ for the Stokes pulse.}
%from Igor File workd0910.pxp
\end{figure}

Similar to Refs.\ \cite{phillipsJMO09, phillipsPRA08}, we extracted the spin wave decoherence time $\tau_s$ by measuring the reduction of the retrieved pulse energies in both the signal and Stokes channels as a function of storage time and fitting to an exponential decay, $e^{-t/\tau_s}$.  Fig.\ \ref{decaygraph} presents a sample measurement of the decay rate at $T=70^\circ$C ($\alpha_0 L=52$), for which we measured $\tau_s=300~\mu$s.  To aid in comparison, we normalized the signal and Stokes retrieval energies to their respective zero-storage-time values, which in this case was $\approx 40\%$ for signal and $\approx 5\%$ for Stokes.  Both the exponential trend and the correspondence between the data %times 
obtained with the signal and Stokes channels are representative of all experimental temperatures, although we found that the spin-wave decoherence rate does have an optical depth dependence.

%%%%%%%%%%%%%%%%%%%%%%%%%%%%%%%%%%%%
%%%%%%%%%%%%%%%%%%%%%%%%%%%%%%%%%%%%
\section{Theoretical description}
%%%%%%%%%%%%%%%%%%%%%%%%%%%%%%%%%%%%
%%%%%%%%%%%%%%%%%%%%%%%%%%%%%%%%%%%%

\subsection{Coupled propagation of signal and Stokes fields in a double-$\Lambda$ system}

\begin{figure*}[hbt]
	\center{\includegraphics[width=1.9\columnwidth]{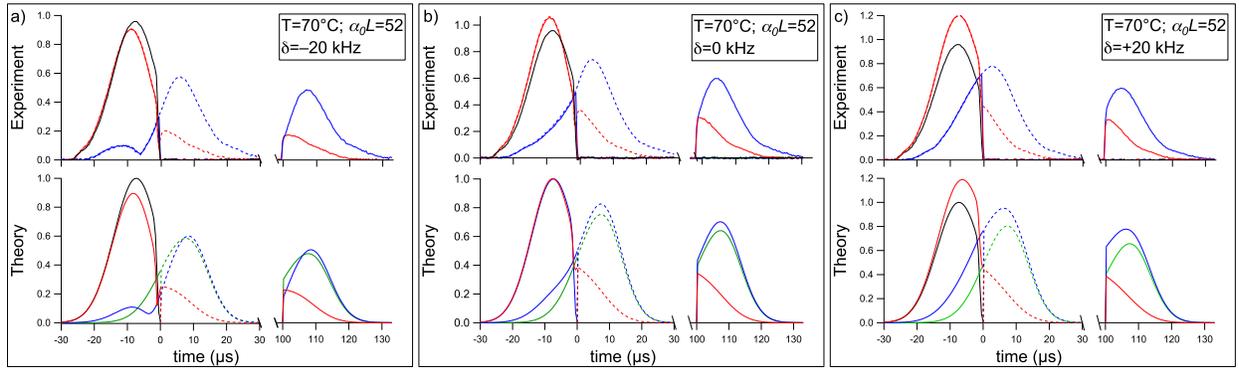}}
	\caption{\label{fig:exppulses} (Color online)  Storage and retrieval of $16~\mu$s-long (FWHM) truncated Gaussian pulses at $T=70^\circ$C ($\alpha_0 L=52$), for a two-photon detuning of (a) $\delta=-20$ kHz, (b) $\delta=0$ kHz, and (c) $\delta=+20$ kHz.  In all plots, the top graphs are experimental data and the bottom graphs are the theoretical predictions from Eqs.\ (\ref{prEq}-\ref{Peq}).  The black curve is a far-detuned reference pulse; the blue (red) traces are the signal (Stokes) pulses.  Dashed (solid) lines correspond to slow (stored) light experiments. In the theory plots, the green curve corresponds to a model consisting only of EIT, to provide contrast with the EIT-FWM model.}
% from Igor File workd0720.pxp
\end{figure*}

In this section, we review the relevant theoretical description of the EIT-FWM process in a double-$\Lambda$ interaction configuration. 
% Previous works have investigated the propagation and interaction of various configurations of two weak light pulses in a double-$\Lambda$ system, which was formed by four real atomic levels \cite{RaczynskiOpComm2002, RZZKPRA2004, CerboneschiPRA, ZimmerOptComm2006, eilamOL}. Additionally, the interaction between EIT and Raman processes has been investigated in a single-$\Lambda$ system in the slow light regime \cite{agarwalFWM, haradaFWM}.  Here, we consider the dynamic interaction of propagating light pulses with a double-$\Lambda$ system, in which the second $\Lambda$ is completed via a virtual transition.
A single %resonant 
$\Lambda$ link consisting of a strong control field (Rabi frequency $\Omega$) and a weak signal field (Rabi frequency $\alpha$) is usually considered in the context of light storage under EIT conditions. In such a single %singular
 $\Lambda$ system under EIT conditions ($\nu-\nu_{\rm c}\simeq \omega_{\rm sg}$, where $\nu$ and $\nu_c$ are signal and control field frequencies, respectively), the control field enables strong coupling between the signal %optical 
 field and the long-lived atomic spin coherence
%such that propagation of a signal pulse in atomic ensemble is accompanied by the through such coherent atomic medium
that is usually described using the formalism of dark-state polaritons %quasi-particle~
\cite{fleischhauer}. Adiabatic variation of the control field power reversibly transfers the signal optical field into the spin coherence, allowing for the realization of a quantum memory.%~\cite{lvovskyNaturePh09,polzikRMP10}.
%Under these conditions   so that any quantum information carried by the signal field can be stored and later retrieved  and the other $\Lambda$ %system is formed by far    In a traditional three-level $\Lambda$ system under conditions of electromagnetically induced transparency a strong %resonant  creates a narrow, symmetric window of transparency is created in the signal field spectrum near the two-photon resonance %($\nu-\nu_{\rm C}\approx \omega_{\rm sg}$).  Correspondingly, the signal field experiences a steep variation in refractive index, leading to a %reduced group velocity (``slow light''), and a pulse delay time of $\tau=(d \gamma)/|\Omega|^2\ll L/c$, where $\gamma$ is the optical %polarization decay rate, $L$ is the length of the interaction region, and $\alpha_0 L$ is the optical depth such that the signal field would be %attenuated by a factor of $\exp{-\alpha_0 L}$ in the absence of the control field.

Achieving sufficiently high memory efficiency, however, requires operation with an atomic ensemble at high optical depth~\cite{gorshkovPRL,novikovaPRLopt,phillipsPRA08}, where the effect on signal field propagation by the off-resonant interaction of the control field on the %$|s\rangle-|e\rangle$ 
$|g\rangle-|e\rangle$
optical transition (Rabi frequency $\Omega'$)  becomes significant and cannot be ignored. In particular, it results in the coherent creation of a Stokes field (frequency $\nu'=\omega_{\rm es}-\Dhf-\delta$, Rabi frequency $\alpha'$) due to four-wave mixing (FWM).  To properly account for this generation~\cite{lukinPRL97, lukinPRLFWM99, CerboneschiPRA, lukinPRL81, RaczynskiOpComm2002, RZZKPRA2004, CerboneschiPRA, ZimmerOptComm2006, eilamOL,agarwalFWM, haradaFWM, phillipsJMO09, howellNatPh09, harrisPRL06, kolchinPRA07, harrisPRL08}, it is necessary to take into account this off-resonant $\Lambda$ link, which is shown in Fig.~\ref{fig:doubleLambda}(a).

The rotating-frame Hamiltonian describing the atomic response to the light fields in such a configuration is:
\begin{equation}\begin{split}
H =& -(\delta -\ds) \ket{s}\bra{s} - (\delta- 2 \ds)\ket{e}\bra{e} \\
&- \left[ \alpha
\ket{e}\bra{g} + \Omega \ket{e} \bra{s} + \frac{\Omega' \alpha'^*}{\Dhf}\ket{s}\bra{g} + {\rm H.c.} \right].
\end{split}\end{equation}
%
%\textcolor{red}{do we want different symbols for $\alpha$, $\alpha'$, since optical depth is $\alpha L$?}
%
Here, the Rabi frequencies of the signal and Stokes fields are $\alpha=E\mu/\hbar$ and $\alpha'=E'\mu'/\hbar$, where $E$ and $E'$ are the slowly-varying envelopes of the signal and Stokes electric fields, correspondingly.  $\mu$ and $\mu'$ are the real dipole matrix elements of the respective transitions.  We define the optical polarization $P(z,t)=\rho_{\rm eg}(z,t)\sqrt{N}$ and the spin coherence $S(z,t)=\rho_{\rm sg}(z,t) \sqrt{N}$, where $\rho_{i j}(z,t)$ is the appropriate slowly-varying position-dependent collective density matrix element and $N$ is the number of atoms in the interaction volume. We use Floquet theory \cite{dreseEPJD99} to adiabatically eliminate the off-resonant interaction via $\Omega'$ and $\alpha'$.  To linear \
%textbf{should we say second order here?} 
order in $1/\Dhf$ and in $\alpha'$, one obtains an effective Rabi frequency $\Omega' \alpha'^*/\Dhf$.  States $\ket{e}$ and $\ket{g}$ acquire small light shifts $\ds=|\Omega'|^2/\Dhf$ and $-\ds$, respectively.  For our Clebsch-Gordon coefficients \cite{phillipsPRA08}, $\Omega'=-\sqrt{3}\Omega$.  In order to match the equations of motion for quantum fields, we define $g\sqrt{N}=\sqrt{\gamma \alpha_0 c/2}$ as the coupling constant between the signal field and the atomic ensemble.  We implicitly assume that the frequencies of the signal and Stokes fields are approximately the same, and we rescale the light field envelopes by defining dimensionless %unitless 
light field envelopes $\e=\frac{\mu}{\hbar g} E$ and $\e'=\frac{\mu}{\hbar g}E'$. %$\e'=\frac{\mu'}{\hbar g}E'$

%Here, the Rabi frequencies of the signal and the Stokes field are $\alpha=\mu\sqrt{\frac{\nu}{2 \epsilon_0 \hbar V}} \e$, $\alpha'=\mu'\sqrt{\frac{\nu'}{2 \epsilon_0 \hbar V}} \e'$, where $\e$ and $\e'$ are the corresponding slowly-varying envelope functions, $\mu$ and $\mu'$ are the real dipole matrix elements of the respective transitions, and $\nu'\approx\nu$ is the light frequency.  We define the optical polarization $P(z,t)=\rho_{\rm eg}(z,t)\sqrt{N}$ and the spin coherence $S(z,t)=\rho_{\rm sg}(z,t) \sqrt{N}$, where $\rho_{i j}(z,t)$ is the appropriate slowly-varying position-dependent collective density matrix element. We use Floquet theory \cite{dreseEPJD99} to adiabatically eliminate the off-resonant interaction via $\Omega'$ and $\alpha'$.  To linear order in $1/\Dhf$ and in $\alpha'$, one obtains an effective Rabi frequency $\Omega' \alpha'^*/\Dhf$.  States $\ket{e}$ and $\ket{g}$ acquire small light shifts $\ds=|\Omega'|^2/\Dhf$ and $-\ds$, respectively.  For our Clebsch-Gordon coefficients \cite{phillipsPRA08}, $\Omega'=-\sqrt{3}\Omega$.

In the dipole approximation, assuming that at all times most of the atoms are in $\ket{g}$, and to linear order in the weak light fields $\e$ and $\e'$, the atomic evolution and light propagation equations read \cite{phillipsJMO09, lukinPRLFWM99, taoFWM, howellNatPh09,lukinPRL81}:
\begin{eqnarray}
\left(\partial_t+c\partial_z\right)\e&=&i g\sqrt{N}P,\label{prEq}\\
\left(\partial_t+c\partial_z\right)\e'^*&=&-i g\sqrt{N}\frac{\Omega}{\Dhf}S,\label{stEq}\\
\partial_t S&=& - \Gamma_0 S+ i \Omega P+i\frac{\Omega}{\Dhf} g \sqrt{N} \e'^*, \label{Seq}\\
\partial_t P&=&-\Gamma P+i \Omega S+i g\sqrt{N}\e, \label{Peq}
\end{eqnarray}
where we have defined $\Gamma_0=\gamma_0-i(\delta-\ds)$ and $\Gamma=\gamma-i(\delta-2\ds)$.  The polarization decay rate $\gamma$ and the spin decay rate $\gamma_0$ have been introduced. %phenomenologically.

Equations (\ref{prEq}-\ref{Peq}) fully describe the propagation of the light fields and the dynamics of the spin wave and of the optical polarization during all stages of light storage.  In the slow light regime, when the control field is constant in time [$\Omega(t)=\Omega$], Eqs.\ (\ref{prEq}-\ref{Peq}) can be Fourier transformed, and Eqs.\ (\ref{prEq}, \ref{stEq}) can be solved analytically \cite{phillipsJMO09, lukinPRLFWM99, taoFWM, howellNatPh09,lukinPRL81}.  In the stored light regime, when the control field intensity is time-dependent, these equations can be solved numerically.
%
%\subsubsection{Correspondence between experiment and theory}
%

Figure \ref{fig:exppulses} displays the results of storage experiments with $16~\mu$s-long truncated Gaussian pulses at $T=70^\circ$C (optical depth $\alpha_0 L=52$) along with the corresponding theoretical predictions, which are obtained by numerically solving Eqs.\ (\ref{prEq}-\ref{Peq}) with the appropriate parameters.  We measured the control field power to be 4.7 mW, and the beam diameter was $2.67$~mm, which corresponded to $\Omega/(2\pi)=9.6$ MHz, and induced a light shift of $\ds=17$~kHz.  The spin wave decay rate was measured to be approximately $300~\mu$s, thus $\gamma_0/(2\pi)\approx270$~Hz.
The results from the slow light experiment (dashed lines) and the stored light experiment (solid lines) are overlaid to facilitate shape comparison.

For a small negative two-photon detuning $\delta=-20$ kHz [Fig.\ \ref{fig:exppulses}(a)], the signal field (in blue) experiences some distortion during propagation [as evidenced by the bumps in the leakage portion of the pulse (when $t<0$), which exits the cell before the control field is extinguished], but the shape of the slow pulse is preserved during the storage process.  
%Likewise, the a fraction of the Stokes field is in the cell during the writing stage; this portion seems to be preserved upon retrieval.  
Likewise, the fraction of the Stokes field that exits the medium at $t > 0$ in the slow light experiment (dashed red) matches the retrieved Stokes field in the stored light experiment (solid red).
There is an excellent agreement between the experimental observations and the numerical model predictions.  The green trace in the theory plot corresponds to standard EIT-based light storage of the signal field, where the FWM process has been artificially turned off.  To compute this trace, the Stokes contribution in Eq.\ \eqref{Seq} is set to zero, and Eqs.\ (\ref{prEq}, \ref{Seq}, \ref{Peq}) are solved numerically.  We include this trace in order to showcase the effects of four-wave mixing on signal pulse shape and delay.

Figure \ref{fig:exppulses}(b) demonstrates the excellent correspondence between experiment and theory for a two-photon detuning of $\delta=0$~kHz.   For this value of $\delta$, the signal pulse is less distorted during propagation, but the pulse shape is still distinct from the bare-EIT model.
Likewise, Fig.\ \ref{fig:exppulses}(c) depicts the results for $\delta=+20$ kHz $\approx \ds$, where the two-photon detuning effectively cancels the light shift during the writing and retrieval stages.  Under this condition, the signal pulse will experience the least amount of distortion due to FWM, since the EIT transmission peak is, at least for a sufficiently narrow pulse bandwidth, symmetric about $\delta=\ds$.  As a result, the dispersion experienced by the pulse is mostly linear.  In all cases, the theoretical model matches the experimental data very well.

The correspondence between slow light pulseshapes (dashed lines) and the shapes of the retrieved pulses (solid lines) illustrates an important result---that when the writing and retrieval control field amplitudes are constant in time, the process of switching the control field off and on has little effect on the signal and Stokes fields, apart from a delay and the spin wave decay during storage time.  In this case, we can further understand the effects of FWM by using the closed form solutions to the Fourier transformed versions of Eqs.\ (\ref{prEq}-\ref{Peq}) \cite{phillipsJMO09, lukinPRLFWM99, taoFWM, howellNatPh09}.  In the Appendix, %of this manuscript, 
we detail the derivation of the following two approximate equations, which intuitively describe the effects of FWM and EIT on pulse propagation for the case $\delta=\ds$.  Although these equations make a set of strong assumptions, including the assumption of an infinitely wide EIT transmission window $\Gamma_E \rightarrow \infty$, they preserve the essential physics in the limit of weak FWM. % In particular, the equations have a simple form, which preserves the essential physics, if we assume that $\Ge\to\infty$.  
Defining $\DR=-\Omega^2/\Dhf$, the equations are
\begin{widetext}
\begin{eqnarray}
\e(L,t) &\approx& \e(0, t - L/v_g) + \DR^2 \int_0^{L/v_g} d t' \e(0, t - t') t' + i \DR \int_0^{L/v_g} d t' \e'^*(0,t-t'), \label{Prapp1}\\
 \e'^*(L,t) &\approx&  \e'^*(0, t) + \DR^2 \int_0^{L/v_g} d t' \e'^*(0, t - t')(L/v_g - t') - i \DR \int_0^{L/v_g} d t' \e(0,t-t'). \label{Stapp1}
\end{eqnarray}
\end{widetext}

These equations clearly show how the effects of FWM grow with optical depth $\alpha_0 L$. The first term on the RHS of Eq.\ (\ref{Prapp1}) describes the delay that the signal field experiences during propagation in an EIT medium, where $v_g = 2 \Omega^2/(\alpha_0 \gamma)$ is the EIT group velocity \cite{fleischhauerRMP05}.  Due to the effects of FWM, the signal field acquires a small in-phase gain of order $\DR^2 (L/v_g)^2 \sim (\alpha_0 L)^2 \gamma^2/\Dhf^2$ from times up to $L/v_g$ earlier.  The farthest away times are weighted more heavily.  Additionally, the signal field acquires an $i$-out-of-phase contribution of order $|\DR| L/v_g \sim \alpha_0 L \gamma/\Dhf$ from the Stokes field up to $L/v_g$ earlier with all times contributing equally.  The Stokes field propagates undistorted and largely undelayed, but gets a small [order $(\alpha_0 L)^2 \gamma^2/\Dhf^2$] in-phase gain from times up to $L/v_g$ earlier, with closest times weighted more heavily, and also an $i$-out-of-phase contribution of order $\alpha_0 L \gamma/\Dhf$ from the signal field, with all times weighted equally. Notice that in both equations, in the regime where the first term on the RHS is large, small in-phase $(\alpha_0 L)^2 \gamma^2/\Dhf^2$ terms and small $i$-out-of-phase $\alpha_0 L \gamma/\Dhf$ terms contribute at the same $(\alpha_0 L)^2 \gamma^2/\Dhf^2$ order to the absolute value of the field (which is what our experiment measures). 

However, the first terms on the RHS are not always dominant. In particular, for $t > 0$, the first term on the RHS of Eq.\ (\ref{Stapp1}) vanishes, in which case $|\e'^*(L,t)|$ is dominated by the third term with a small correction from the second term. This means, as we will confirm experimentally, that the retrieved Stokes field is largely determined by the input signal, and not by the input Stokes field. Similarly, if EIT group delay is comparable to the signal pulse duration, then, for $t < 0$, the RHS of Eq.\ (\ref{Prapp1}) is small and $|\e(L,t)|$ is significantly affected by the third term. This means, as we will confirm experimentally, that the signal pulse leakage can be strongly affected by the Stokes input, in contrast to the retrieved signal pulse, which is affected by the Stokes input only weakly. Eqs.\ (\ref{Prapp1}, \ref{Stapp1}) also show that the perturbative treatment of the effects of FWM, employed to derive them, breaks down when $|\DR L/v_g| \gtrsim 1$, \textit{i.e.}, when the optical depth is $\alpha_0 L \gtrsim 2 \Dhf/\gamma \approx 100$.  

\begin{figure}[tb]
\center{\includegraphics[width=\columnwidth]{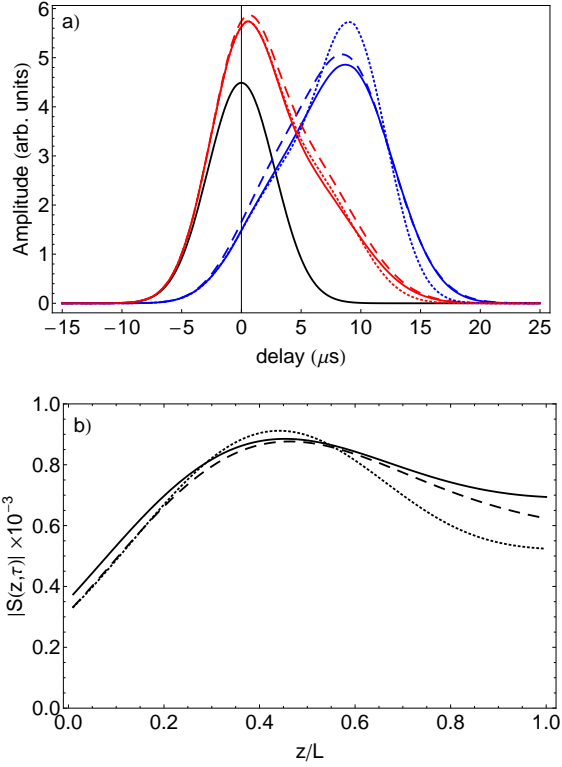}}
	\caption{\label{fig:pulsesandspinwave}(Color online) (a) Results of a numerical investigation of slow light with a $6.6~\mu$s-long pulse (reference in black), with $\Omega/(2\pi)=10$ MHz and $\alpha_0 L=80$, so that the bandwidth of the pulse $\Delta\omega=0.1\Ge$.  Blue traces are the signal field; red traces are the Stokes field.  Solid lines are the result of numerically solving Eqs.\ (\ref{prEq}-\ref{Peq}).  Dotted lines are the result of the infinite-$\Ge$ approximation in Eqs.\ (\ref{Prapp1}-\ref{Stapp1}).  Dashed lines correspond to results obtained using numerical integration of expressions in Eqs.\ (\ref{eEqwithf}-\ref{ep}, \ref{f1}-\ref{g3}) in the Appendix.  (b) The spin waves created at  the time corresponding to a $5~\mu$s delay in Fig.\ \ref{fig:pulsesandspinwave}(a). %when portions of the Stokes and signal pulses overlap in the cell.  
The solid black line is the result of numerically solving Eqs.\ (\ref{prEq}-\ref{Peq}).  Dotted lines are the results from Eq.\ \eqref{Sapp1}.  Dashed lines are the results from Eqs.\ (\ref{Spinwithh}, \ref{happrox}).}
% from Igor File d01052011.pxp
\end{figure}

To test the validity of Eqs.\ (\ref{Prapp1}, \ref{Stapp1}), in Fig.\ \ref{fig:pulsesandspinwave}(a), we compare the solutions obtained by numerically solving Eqs.\ (\ref{prEq}-\ref{Peq}) (solid lines) to the predictions of Eqs.\ (\ref{Prapp1}, \ref{Stapp1}) (dotted lines).  In the dashed traces, we include the results of a useful intermediate approximation, which does not assume infinite $\Ge$ and is described by Eqs.\ (\ref{eEqwithf}-\ref{ep}, \ref{f1}-\ref{g3}) %(\ref{eEqwithf}-\ref{g3expr}) 
in the Appendix.  For these plots, $\Omega/(2\pi)=10$ MHz and $\alpha_0 L=80$; the pulse bandwidth was $\Delta\omega=0.1\Ge$.  From the excellent 
correspondence between theoretical models, it is evident that the approximations made in deriving Eqs.\ (\ref{Prapp1}, \ref{Stapp1}) are valid.
%preserve the essential physics. 

\subsection{Effect of four-wave mixing on the spin wave}

%\textcolor{red}{Here, we describe the approximations and derive the equation of propagation for F.  Describe what the equation predicts: \\
%1) we should modify the storage equation. \\
%1b) These equations look very similar to those of regular storage, but with F instead of $\e$.  \\
%2) Get shape-preserving propagation of F at low optical depths. \\
%2b) Can we derive the optical depth where the stokes term matters?\\
%3) The stokes term results in gain of the stokes component of F.\\
%4) Storage: S(z) is mostly signal.  Since F is largely (98\%) probe, filtering the Stokes seed does not really change S.\\
%5) Retrieval: S(z) is mapped back into F, whose stokes component acts like a seed for FWM.  So, we get the same Stokes out regardless of input.}

While the solutions of Eqs.\ (\ref{prEq}-\ref{Peq}) accurately describe the evolution of light pulses and atomic variables under slow light and storage conditions, %they do not
we have not yet used them to elucidate the role that the Stokes field plays in the creation of the spin coherence.  Specifically, it is not yet clear whether the quantum memory description based on the dark state polariton principle \cite{fleischhauer} is valid under EIT-FWM conditions.  In what follows, we develop a more transparent description of light storage in a double-$\Lambda$ system and show that in this case, the spin wave is determined by a particular combination of signal and Stokes fields.

We obtain this result by adiabatically eliminating the optical polarization $P(z,t)$.  We set the time derivative to zero in Eq.\ \eqref{Peq} and find
\begin{equation}
P(z,t)\approx i\frac{\Omega}{\Gamma} S(z,t)+i \frac{g\sqrt{N}}{\Gamma}\e(z,t).
\label{Papprox}
\end{equation}
Inserting Eq.~\eqref{Papprox} into Eq.~\eqref{Seq}, we obtain the following equation for time evolution of the spin wave $S(z,t)$:
\begin{equation}
\partial_t S(z,t)=-\left(\Gamma_0+\frac{\Omega^2}{\Gamma}\right) S(z,t) - g\sqrt{N} \frac{\Omega}{\Gamma} \mathcal{F}(z,t).
\label{newSEq}
\end{equation}
It is easy to see that the spin wave depends only on a combination $\mathcal{F}(z,t)$ of signal and Stokes optical fields, defined as
\begin{equation}
\mathcal{F}(z,t)=\e(z,t)-i\frac{\Gamma}{\Dhf} \e'^*(z,t).
\label{eq:Fexpression}
\end{equation}
%This combination $\mathcal{F}(z,t)$ describes a combination of signal and Stokes fields that determines the spin coherence. 
Eq.~\eqref{newSEq} is analogous to the spin wave expression obtained through a similar treatment of a standard three-level light storage model \cite{fleischhauer},
\begin{equation}
\partial_t S(z,t)=-\left(\Gamma_0+\frac{\Omega^2}{\Gamma}\right) S(z,t) -g \sqrt{N} \frac{\Omega}{\Gamma} \e(z,t),
\label{3levSeq}
\end{equation}
but with one modification---the single light field (signal) is now replaced by a combined signal-Stokes field $\mathcal{F}$.  Thus, one might expect that it should be possible to store information about this joint mode in the spin coherence.  However, only a small  $(\alpha_0 L)^2 \gamma^2/\Dhf^2$ fraction 
 of the Stokes field [the second term on the RHS of Eq.\ \eqref{Stapp1}] exits the medium at $t > 0$ after the input Stokes has been turned off,
% does experience a reduced group velocity during propagation---an effect proportional to $(\alpha_0 L)^2 \gamma^2/\Dhf^2$ (\textbf{is this correct?}) [see Eq.\ \eqref{Stapp1}]--- 
while the signal pulse is delayed in its entirety [the first term on the RHS of Eq.\ \eqref{Prapp1}]. 
%velocity is  much lower.  
As a result, in contrast to the information encoded in the signal field, most of the information encoded in the Stokes field is lost to leakage, which leaves the interaction region before the control field is shut off.  

%At low optical depths, the same spin wave is created, regardless of input Stokes amplitude.
%Notice, however, that this storage does not imply independent preservation of either signal or Stokes fields individually: both of them can be modified during the propagation as long as their $F$ combination stays the same. This remains true even in case of no seeded input Stokes field, considered in Ref.~\cite{howellNatPh09}.  The observation that spin decay rates are the same with both Stokes and signal pulse measurements (see Fig.\ \ref{decaygraph}) is consistent with this interpretation of storage.
The similarity between Eq.\ \eqref{newSEq} and Eq.\ \eqref{3levSeq} motivates a more detailed comparison of our EIT-FWM system with the traditional EIT configuration.
%an investigation of the dark state polariton interpretation of storage. (AG: I don't want to mention the dark state polariton here, so that people don't start thinking of F as being similar to the dark state polariton, because it is really similar to E but not to the dark state poalriton)
The propagation equation for $\mathcal{F}$ is easily obtained from the appropriate combination of Eqs.\ (\ref{prEq}, \ref{stEq}):
\begin{eqnarray}
\left(\partial_t+c\partial_z\right)\mathcal{F}=-\frac{g^2 N}{\Gamma} \mathcal{F} - \Omega \frac{g\sqrt{N}}{\Gamma} S- i \frac{g^2 N}{\Dhf} \e'^*.
\label{FpropEq}
\end{eqnarray}
This equation is also similar to the signal propagation expression in the classic stored light model \cite{fleischhauer},
\begin{equation}
\left(\partial_t+c\partial_z\right)\e=-\frac{g^2 N}{\Gamma} \e - \Omega \frac{g\sqrt{N}}{\Gamma} S,
\end{equation}
except for the optical-depth-dependent Stokes term, which describes the generation of signal from Stokes during propagation through a sufficiently optically-thick medium.
%\textcolor{red}{add brief discussion about similarity with dark state polariton at small optical depth, estimate of maximum od at which this approximation is still valid, and how it gets broken because of Stokes gain. We had these discussions a few times, so you probably have all calculations.}

\begin{figure}[t!]
	\center{\includegraphics[width=0.48\textwidth]{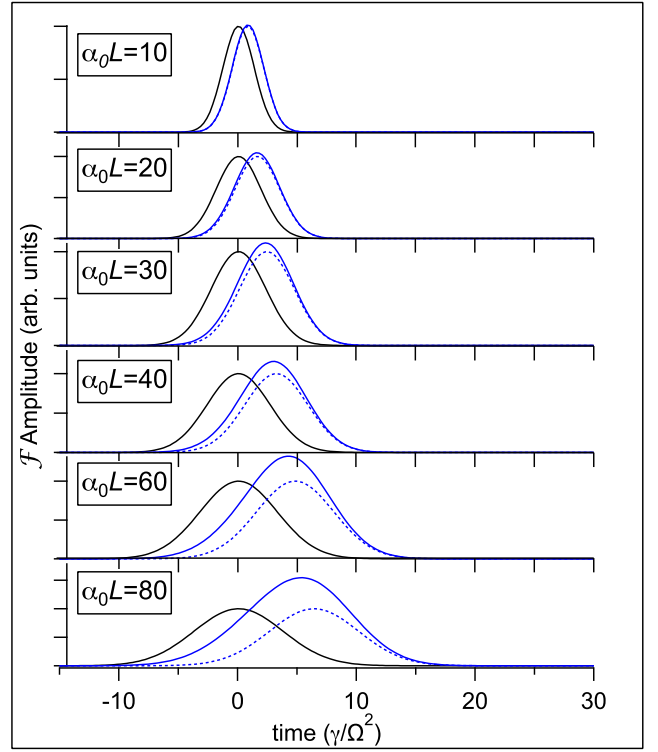}}
	\caption{\label{fig:FvsOD}(Color online) Results from numerical evaluation of Eqs.\ (\ref{stEq}, \ref{newSEq}, \ref{Fsloweq}) (solid blue lines) and of the homogeneous version of Eq.\ (\ref{Fsloweq}) (dashed blue lines) for a range of optical depths $\alpha_0 L$, as indicated in the legends.  The bandwidth of each input pulse was $\Delta\omega=0.05\Ge=\Omega^2/(20\sqrt{\alpha_0 L/2}\gamma)$.}
% from MA file FvsOpticalDepth, and similar Igor File.
\end{figure}

%If the control field intensity varies adiabatically, we can iterate Eq.~\eqref{newSEq} to obtain an approximate expression for $S(z,t)$:
%%
%\begin{equation}
%S(z,t) \approx \frac{g\sqrt{N} \Gamma}{\Omega^3} \partial_t \mathcal{F}(z,t) - \frac{g\sqrt{N}}{\Omega} \mathcal{F}(z,t),
%\label{eq:SmappedtoF}
%\end{equation}
%where we have implemented the EIT condition $\Omega^2\gg|\Gamma \Gamma_0|$.  
When the two-photon detuning is chosen such that the light shift is canceled (\textit{i.e.,} $\delta=\ds$), the propagation equation becomes, to $\mathcal{O}(1/\Dhf),$
\begin{equation}
\left[\partial_t+c \cos ^2\theta(t)\partial_z\right]\mathcal{F}(z,t)\approx i \DR \e'^*(z,t)
\label{Fsloweq}
\end{equation}
with the angle $\theta(t)$ given by $\tan^2\theta(t)=\frac{g^2N}{\Omega^2(t)}$. %, and we have defined $\DR=-\Omega^2/\Delta$.  

Analysis of above equations demonstrates two regimes for light storage under EIT-FWM conditions. At low optical depths, the contribution of the Stokes field on the RHS of Eqs. (\ref{FpropEq}, \ref{Fsloweq}) is negligible.  In this case, the equations for the joint field $\mathcal{F}$ and spin wave $S$ become identical to those for %a tranditional dark state polariton (AG: I don't want to mention the dark state polariton here to avoid confusion of F with the dark state polariton
$\e$ and $S$ in the regular EIT configuration. For example, if we replace the RHS of  Eq.\ \eqref{Fsloweq} with zero, it would describe the propagation of $\mathcal{F}$ without distortion at a reduced group velocity $v_g=c \cos^2 \theta\approx \frac{2 \Omega^2}{\alpha_0 \gamma}$.  However, at low optical depths and $t > 0$ (after the input Stokes pulse has been turned off), the contribution of the Stokes field into $\mathcal{F}$ is also negligible: it is small not only because of the small factor $\Gamma/\Dhf$ in Eq.~(\ref{eq:Fexpression}) but also because $\e'(z,t)$ itself is small [since the first term on the RHS of Eq.\ (\ref{Stapp1}), generalized to arbitrary $z$, vanishes for $t > 0$].
%not only by $\Gamma/\Dhf$ from Eq.~(\ref{eq:Fexpression}) but also by $(\alpha_0 L)^2 \gamma^2/\Delta^2$ from the second term on the RHS of Eq.\ (\ref{Stapp1}). 
Thus, signal field propagation can be analyzed using a three-level single $\Lambda$, even though the Stokes field %by itself 
can be significantly affected by control and signal fields, as is evident from the dominance of the last term on the RHS of Eq.\ (\ref{Stapp1}) for $t > 0$.  

However, at higher optical depths, the term on the RHS of Eq.~\eqref{Fsloweq} becomes significant.  Specifically, this term results in gain or loss of the signal field due to the Stokes field. 
%This results in gain of the imaginary part of $F$ (\textit{e.g.,} Stokes component) during propagation.  
The dashed blue lines in Fig.\ \ref{fig:FvsOD} depict the results of numerical calculations of the homogeneous form of Eq.\ \eqref{Fsloweq}.  Solid blue lines show the results of the numerical evaluation of the full form of Eq.\ \eqref{Fsloweq} with Eqs.\ (\ref{stEq}, \ref{newSEq}).  For these calculations, $\Omega/(2\pi)= 8$ MHz, $\gamma/(2\pi)=150$ MHz, and the signal pulse was chosen so that its bandwidth, $\Delta\omega=0.05\Ge$.  It is evident from this graph that the Stokes contribution is not negligible %ignorable 
for optical depths %above 
$\alpha_0 L\gtrsim 25$, when the simple slow propagation of $\mathcal{F}$ %dark state polariton interpretation 
breaks down due to the Stokes term on the RHS of Eq.\ (\ref{Fsloweq}).

\begin{figure}[t]
	\center{\includegraphics[width=\columnwidth]{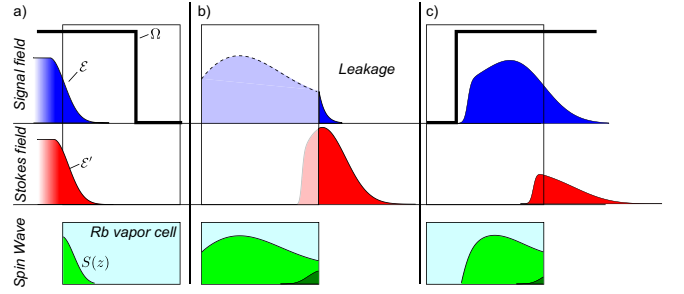}}
	\caption{(Color online) An illustration of the modified storage description.  (a)   During the writing stage, the input signal field $\e$ (\textit{Top}) and Stokes field $\e'$ (\textit{Middle}) propagate at different
	%(different) reduced 
	group velocities through the atomic medium, creating a spin wave (\textit{Bottom}).  (b)  During the storage stage, the control field is turned off and no light fields are present.  Some portion of the signal field has propagated through the cell and leaks out before the control field is extinguished.  At the same time, most of the information in the Stokes field is lost in the leakage, since, in the regime $(\alpha_0 L) \gamma/\Delta \ll 1$, the propagation of the Stokes field is affected by the atoms only weakly   [see Eq.\ (\ref{Stapp1})].
	 %the bulk of the Stokes field experiences no group delay 
	% [see the first term in Eq.\ (\ref{Stapp1})]. 
	%Stokes pulse typically only experiences a modest change in group velocity.  
	The spin wave is preserved during storage.  (c)  During retrieval, the control field is turned on, releasing the spin wave into both the signal and Stokes fields, which exit the vapor cell.
	\label{storagecartoon}}
\end{figure}

%
%\begin{figure}[t!]
%\psfrag{Amp}{$\mathcal{F}$ Amplitude (arb. units)}
%\center{\includegraphics[width=0.48\textwidth]{figures/FvsOD2b.eps}}
%	\caption{\label{fig:FvsODb} Results from numerical evaluation of Eqs.\ \eqref{Fsloweq} (solid blue lines) and its homogeneous version (dashed blue lines) for a range of optical depths.  The bandwidth of each input pulse was $\Ge/20=\Omega^2/(20\sqrt{\alpha_0 L/2}\gamma)$.}
%% from MA file FvsOpticalDepth, and similar Igor File.
%\end{figure}

As shown in the Appendix, the same approximations that lead to Eqs.\ (\ref{Prapp1},\ref{Stapp1}) give the following expression for the spin wave $S(z,t)$ in the limit when $\Ge\to\infty$:
\begin{equation}\begin{split}
S(z,t)\approx-\frac{g\sqrt{N}}{\Omega}&\left[\e(0,t-\frac{z}{v_g}) + \DR^2\int_0^{\frac{z}{v_g}} dt' \e(0,t-t') t' \right. \\
&\left.+i\DR\int_0^{\frac{z}{v_g}} dt' \e'^*(0,t-t')\right].
\label{Sapp1}
\end{split}
\end{equation}

In Fig.\ \ref{fig:pulsesandspinwave}(b), we compare the shape of the spin wave that is obtained by numerically solving Eqs.\ (\ref{prEq}-\ref{Peq}) (solid lines) to the predictions of Eq.\ \eqref{Sapp1} (dotted lines).  As in Fig.\ \ref{fig:pulsesandspinwave}(a), we also include the predictions of an intermediate approximation, which does not assume an infinite $\Ge$ and is described in Eqs.\ (\ref{Spinwithh}, \ref{happrox}) in the Appendix. The reasonable agreement between the three curves in Fig.\ \ref{fig:pulsesandspinwave}(b) implies that Eq.\ (\ref{Sapp1}) does indeed contain the essential physics. In particular, under this approximation, the spin wave is proportional to the signal field only, as in a traditional three-level single-$\Lambda$ EIT system [see Eq.\ (\ref{happrox}) in the Appendix], 
\begin{equation}
S(z,t) \approx - \frac{g \sqrt{N}}{\Omega} \e(z,t). \label{sapp}
\end{equation}
%$ [see Eq.\ (\ref{sapp})], 
Moreover, under this approximation, $\e(z,t)$ [and hence $S(z,t)$] is mostly determined by the usual slowed-down version of the input signal [first term in the square brackets in Eq.\ (\ref{Sapp1})] with small corrections of order $|\DR| L/v_g \sim \alpha_0 L \gamma/\Dhf$  [third term in the square brackets in Eq.\ (\ref{Sapp1})] and $(\alpha_0 L)^2 \gamma^2/\Dhf^2$ [second term in the square brackets in Eq.\ (\ref{Sapp1})].

%It is important to notice that Eq.\ \eqref{Sapp1} is identical to the signal field expression in Eq.\ \eqref{Prapp1}, except for the scaling factor $-g\sqrt{N}/\Omega$.  In other words, up to order $1/\Dhf^2$, we find that, as in the three-level EIT scheme, the spin wave, $S(z,t)\propto \e(z,t)$, although the propagation dynamics of the light field are modified by the FWM.

Fig.\ \ref{storagecartoon} illustrates an intuitive way to understand storage under EIT-FWM conditions.  At the beginning of the writing stage, shown in Fig.\ \ref{storagecartoon}(a), the control field (in black) prepares the atoms and causes the input signal field $\e$ (in blue, \textit{top}) to propagate at a reduced group velocity.  The Stokes pulse (in red, \textit{middle}) enters the cell and is not completely extinguished inside the medium even after the reference pulse would have left the medium. 
%in many cases can experience a slight reduction of group velocity. 
 A collective spin coherence is created in the atomic vapor cell (in green, \textit{bottom}).  As the pulses propagate through the atomic medium, as shown in column (b), they experience mutual interference effects and may become distorted.  The spin wave propagates along with the signal field.  The contributions to the spin wave are determined by the joint mode $\mathcal{F}$, and we can discern between the contributions to the spin wave from the signal field (shown in light green) from those of the Stokes field (shown in dark green)  [see Eq.\ \eqref{Sapp1}].  In the regime of weak FWM, the propagation of the Stokes field is only weakly affected by the atoms [see Eq.\ (\ref{Stapp1})], so that much of the Stokes field
 %The bulk of the Stokes field [the first term in Eq.\ (\ref{Stapp1})] propagates at the speed of light and
 %Stokes field does not slow down as much as the signal field, much of it 
 leaves the end of the vapor cell as leakage.  Any information contained in this leaked field is lost for the storage process, which commences when the control field is shut down.  After some time [column (c)], the control field is turned on, and the spin wave is released into both the signal and Stokes fields.  It is important to note that, since the joint mode $\mathcal{F}$ is not a normal mode, the proportion of Stokes to signal is not fixed.

In the regime of weak FWM ($\alpha_0 L \leq 25$), the joint mode $\mathcal{F}(z,t)$ is determined mostly by the input signal field $\e$.  Thus, the propagation dynamics experienced by the signal pulse will be only slightly sensitive to the amplitude of the seeded Stokes pulse [of order $\alpha_0 L \gamma/\Dhf$, see the last term in Eq.\ \eqref{Prapp1}], and consequently the spin wave created will have the same weak dependence on the seeded Stokes field [see the last term in Eq.\ \eqref{Sapp1}].  
%On the other hand, the Stokes field propagation dynamics depend more strongly on the amplitude of the Stokes seed [see Eq.\ \eqref{Stapp1}].  
Since the spin wave is only weakly dependent on the input Stokes field, it is possible to create approximately the same spin wave for different input combinations of signal and Stokes fields.  The retrieval from the spin wave into the light fields %should 
will consequently have 
this same weak dependence on the input Stokes field. 

Eqs.\ (\ref{Prapp1},\ref{Stapp1}) support this conclusion. Specifically, the amplitude of the retrieved signal field [Eq.\ (\ref{Prapp1})] is determined primarily by the input signal field (the first term on the RHS) with a small $(\alpha_0 L \gamma/\Dhf)^2$ correction from the input Stokes field (the third term on the RHS). Similarly, since the first term on the RHS of Eq.\ (\ref{Stapp1}) vanishes for $t > 0$, the retrieved Stokes field is also determined primarily by the input signal (the third term on the RHS) with a small  $(\alpha_0 L \gamma/\Dhf)^2$ correction from the input Stokes (the second term on the RHS). At the same time, signal and Stokes outputs are more strongly affected by the Stokes input for $t < 0$ (leaked pulses) than for $t > 0$ (retrieved pulses). This statement is obvious for the Stokes field, since the first term on the RHS of Eq.\ (\ref{Stapp1}) does not vanish for $t < 0$.  The reason this statement holds for the output signal %The effect of input Stokes on the signal output is larger for $t < 0$ 
is that the first term on the RHS of Eq.\ (\ref{Prapp1}) is smaller for $t<0$ than for $t > 0$ for a sufficiently large group delay, while the third term on the RHS of Eq.\ (\ref{Prapp1}) is larger for $t<0$ than for $t > 0$ since $\e'(0,t-t')$ vanishes for $t-t' > 0$.

 %[the third term on the RHS of Eq.\ (\ref{Prapp1}) and the second term on the RHS of Eq.\ (\ref{Stapp1})]. 
% In fact, Eq.\ (\ref{Stapp1}) shows that the retrieved Stokes field ($t > 0$) is determined primarily 

\begin{figure}[tbp]
	\center{\includegraphics[width=\columnwidth]{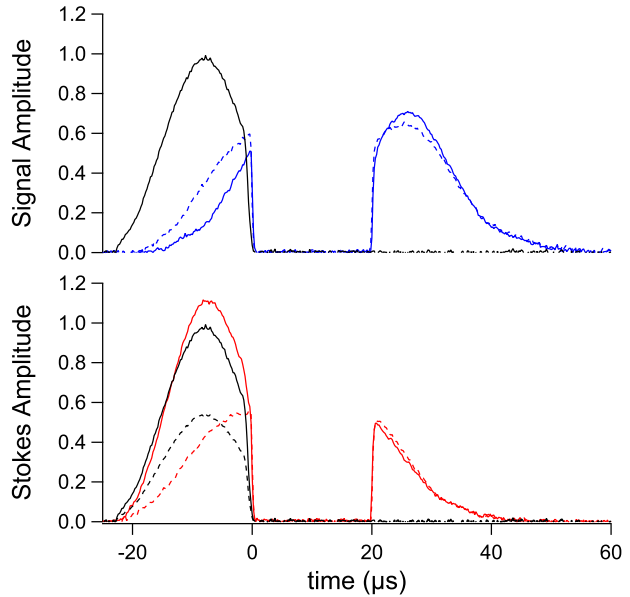}}
	\caption{\label{fig:FvsNF}(Color online) Storage of a $15~\mu$s (FWHM) truncated Gaussian pulse at $T=70^\circ$C ($\alpha_0 L=52$) under different Stokes seeding conditions.  Solid lines depict storage when the Stokes seed amplitude is the same as the signal amplitude.  The dashed lines correspond to the case of a reduced input Stokes field.  The black traces show reference (input) pulses, and the dashed black trace in the bottom plot illustrates the reduced Stokes seed amplitude.}	
%	Filtered vs. Non-Filtered.  $\Omega=6$ MHz, $\ds=20$ kHz, pow=9.7x1.8.  diam=5.35mm
\end{figure}

In Fig.\ \ref{fig:FvsNF}, we display the results of an experiment designed to test these conclusions. %is interpretation.  
The top graph [Fig.\ \ref{fig:FvsNF}(a)] depicts the storage of a $15~\mu$s-long (FWHM) truncated Gaussian signal field at $T=70^\circ$C ($\alpha_0 L=52$).  The solid trace corresponds to approximately equal amplitudes of input signal and Stokes optical pulses, while  the dashed traces correspond to a reduced initial Stokes amplitude $\e'^*(0,t)=-0.55\e(0,t)$.
%$f=\sqrt{0.3}$, when has been attenuated.

Notice the difference in the leakage portion ($t < 0$) of both the signal pulse and the Stokes pulse as we go from solid curves to dashed curves, which exemplifies that both signal and Stokes outputs for $t < 0$ do depend strongly
%how the propagation dynamics of each pulse depends %strongly 
on the amplitude of the seeded Stokes pulse, as we have explained theoretically above and as we have reported previously \cite{phillipsJMO09}.  %On the contrary, 
At the same time, the retrieved ($t > 0$) Stokes and signal pulses are both almost independent 
of the %relative 
amplitude of the seeded Stokes field, which is consistent with the theoretical explanation above.
%Eq.\ (\ref{Sapp1}) and the interpretation that the spin wave is weakly affected by the seeded Stokes amplitude.  
We repeated similar measurements many times under a wide range of experimental conditions and found the retrieved pulses to be weakly affected by the seeded Stokes amplitude as long as the input signal field is comparable to or stronger than the input Stokes field.
%
%Notice that despite an obvious difference between these two cases in portion of the signal pulse that leaked from the cell before control field was turned off ($t<0$), the retrieved pulses are similar.  Likewise, the Stokes leakage [Fig~\ref{fig:FvsNF}(b)] is also different in each case, but the retrieval is very nearly identical. 
%These graphs emphasize an important difference between slow and stored light pulse propagations under EIT-FWM conditions.   As we discussed in a earlier work~\cite{phillipsJMO09}, in the slow light regime, the shapes of both the signal and Stokes optical fields that exit the interaction region (as seen in the leakage portion of the pulse for $t<0$) are the results of a standard interference phenomenon.  This interpretation holds in the present case, when a strong effect is expected even if the spin coherence is solely determined by signal and control fields% and does not depend on the Stokes field at all.   
%It is expected that the amplitude of either signal or Stokes field should depend on the initial amplitude of the latter. 
%However, during light storage, all optical fields vanish once the control field is turned off, and only the information stored in the spin wave can be recreated.  Therefore, we expect only a weak modification of the retrieved pulses for different Stokes inputs since the joint mode $\mathcal{F}$ and the spin wave have only weak dependence on the input Stokes field, as we see in the experiments.

\section{Optical depth dependence of the Stokes field}
%
%\begin{figure}[t]
%	\center{\includegraphics[width=0.5\textwidth]{figures/StokesVsOD.eps}}
%	\caption{\label{fig:stokesvsOD} Stokes behavior for increasing optical depths.  The black trace is a reference pulse; the blue (red) trace is the probe (Stokes) pulse.  (a) Storage and retrieval of a $6~\mu$s-long (FWHM) Gaussian pulse at $T=50$, which corresponds to an optical depth of $\alpha_0 L=10.$  Here, $\Omega=(2\pi)~8.3$ MHz.  (b)  Pulse duration is $6~\mu$s, $\Omega=(2\pi)~7.1$ MHz.   (c) Pulse duration is $20~\mu$s, $\Omega=(2\pi)~12.7$ MHz.  (d)  Pulse duration is $20~\mu$s, $\Omega=(2\pi)~7.8$ MHz. \textcolor{red}{Include $^\circ$C and `slow/stored light' labels.}}
%% from Igor File d01052011.pxp
%\end{figure}

%  \textcolor{red}{maybe say more about this graph?}

\begin{figure*}[t]
	\center{\includegraphics[width=\textwidth]{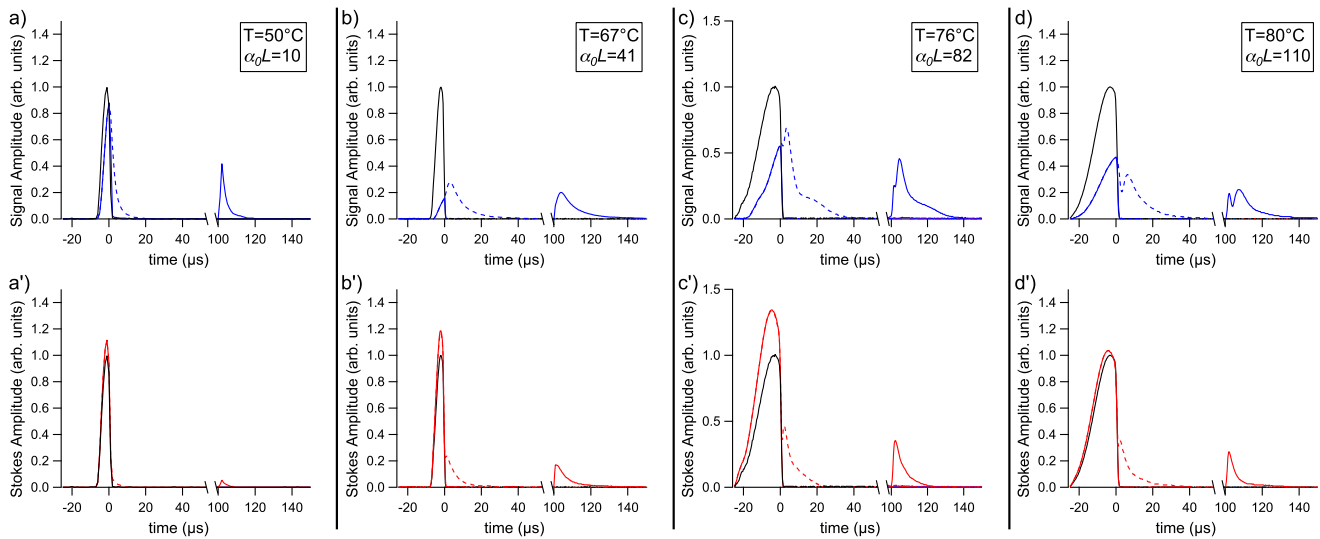}}
	\caption{\label{fig:stokesvsOD}(Color online) Stokes behavior for increasing optical depths.  For all cases, the two-photon detuning $\delta=0$.  The black trace is a reference pulse; the blue (red) trace is the signal (Stokes) pulse.  (a, a$'$) Storage and retrieval of a $6~\mu$s-long (FWHM) truncated Gaussian pulse at $T=50^{\circ}$, which corresponds to an optical depth of $\alpha_0 L=10.$  Here, $\Omega/(2\pi)=8.3$ MHz.  (b, b$'$)  $T=67^{\circ} (\alpha_0 L=41)$, pulse duration is $6~\mu$s, $\Omega/(2\pi)=7.1$ MHz.   (c, c$'$) $T=76^{\circ} (\alpha_0 L=82)$, pulse duration is $20~\mu$s, $\Omega/(2\pi)=12.7$ MHz.  (d, d$'$)  $T=80^{\circ} (\alpha_0 L=110)$, pulse duration is $20~\mu$s, $\Omega/(2\pi)=7.8$ MHz.}
% from Igor File d01052011.pxp
\end{figure*}

In this section, we present the results of storage experiments at increasing optical depths.  

Figure \ref{fig:stokesvsOD} depicts the evolution of the Stokes and signal fields under storage conditions as optical depth increases.  The general features of these results are well-explained by the simple signal and Stokes expressions in Eqs.\ (\ref{Prapp1}-\ref{Stapp1}), as described below.  In Figs.\ \ref{fig:stokesvsOD}(a, a$'$), we show the results of slow light and stored light experiments using a $6~\mu$s-long (FWHM) truncated Gaussian pulse at a temperature of $T=50^\circ$C, which corresponds to an optical depth of $\alpha_0 L=10$.  The signal pulse [blue trace in Fig.\ \ref{fig:stokesvsOD}(a)] experiences a reduction in group velocity during propagation, as seen by comparing the dashed trace (slow light) to the black trace, which is a far-detuned reference pulse.  The Stokes pulse [red trace in Fig.\ \ref{fig:stokesvsOD}(a$'$)] closely mimics the far-detuned reference pulse, indicating that four-wave mixing is not a dominant process at this optical depth.  In a separate run, we investigate storage of these pulses by turning off the control field for $100~\mu$s.  Upon retrieval, the signal field [solid blue trace in Fig.\ \ref{fig:stokesvsOD}(a)] is recovered with a modest reduction in amplitude due to spin wave decay during the storage time, but its shape is preserved.  Additionally, we retrieve a small pulse on the Stokes channel [solid red trace in Fig.\ \ref{fig:stokesvsOD}(a$'$)].
%, whose shape appears to be an attenuated copy of the retrieved signal pulse; AG: do we have any explanation for why this would be the case?  I would say from the last term in Eq. (7) that this is NOT the case - the output Stokes is a smeared out version of the input signal 
Eq.\ \eqref{Prapp1} predicts that at low optical depths [$(\alpha_0 L \gamma/\Dhf) \ll 1$] %
%[\textit{i.e.,} ignoring the second term $\propto (\alpha_0 L \gamma/\Dhf) \ll 1$], 
the retrieved signal pulse will be a delayed version of the input pulse (if one accounts for the storage time), but with a slight modification due to the Stokes field [the last term in Eq.\ (\ref{Prapp1})].  Likewise, the Stokes field will be mostly unaffected by the atoms [the first term on the RHS of Eq.\ (\ref{Stapp1})], so most of it will leak out (see $t<0$). However, a small Stokes pulse $\propto (\alpha_0 L/\Dhf)$ generated from the input Signal [the last term in Eq.\ (\ref{Stapp1})] will be retrieved. %, as expected from Eq.\ \eqref{Stapp1}.

Figures \ref{fig:stokesvsOD}(b, b$'$) show the results of similar experiments at $T=67^\circ$C, corresponding to an optical depth of $\alpha_0 L=41$.  Here, the signal shape is again preserved by the storage process.  The four-wave mixing effects are exhibited by the Stokes gain in the leakage portion of the pulse [see Fig.\ \ref{fig:stokesvsOD}(b$'$)], which leaves the interaction region 
%(and hence reaccelerates to $v_g\approx c$) 
before the storage stage occurs. This gain is described by the last two terms in Eq.\ (\ref{Stapp1}).
%The Stokes field's shape at the retrieval stage mimics its shape during the slow light process, and the Stokes shape appears to be similar, but not identical, to the signal shape.  AG: again, I think our theoretical interpretation does not support a claim of this sort; also see, for example, Figure 1(b), where the two shapes are not really similar, I would say.
At this increased optical depth, the last term in Eq.\ \eqref{Stapp1} also predicts an increased Stokes output for $t > 0$. 
%predicts additional Stokes gain upon retrieval.  
The effects of FWM are also apparent in the distortion that the signal field experiences during propagation.

Figures \ref{fig:stokesvsOD}(c, c$'$) depict the storage experiments at $T=76$ ($\alpha_0 L=82$).  We used a longer pulse (FWHM of $20~\mu$s).  At this optical depth, the Stokes pulse experiences more gain during propagation.  Storage and retrieval, however, still preserve the shapes of both the signal and the Stokes pulse. %, which are clearly distinct from one another.  
Again, it is clear that the Stokes field gain predicted by Eq.\ \eqref{Stapp1} becomes more apparent at higher optical depths.  We also see that, at $\alpha_0 L=110$ [column (d)], the Stokes field amplitude is smaller than at $\alpha_0 L=82$ [column (c)]. 
This effect is most likely %either due to the break down of the perturbative expansion used to derive breaks down at $\alpha_0 L \approx 100$ or to 
due to the absorption of the control field by unprepared atoms that enter the interaction region during the waiting time. We also note that, at $\alpha_0 L \approx 100$, the perturbative expansion used to derive Eqs.\ (\ref{Prapp1},\ref{Stapp1}) breaks down. 

\section{Conclusion}

We have studied the phenomenon of stored light under conditions of electromagnetically induced transparency (EIT) and four-wave mixing (FWM) in an ensemble of hot Rb atoms.  In particular, we have investigated the prospect of simultaneously storing both a signal and a Stokes pulse in a single atomic coherence, and have shown that independent storage of two modes is not possible. The reason is that most of the Stokes pulse leaks out of the medium during the writing stage, so that during retrieval both output fields are determined primarily by the input signal field and depend on the input Stokes field only very weakly. We presented a theoretical model based on a simple double-$\Lambda$ system, which agreed very well with experimental observations. This model allowed us to derive a simple relationship between input and output fields, which explained the above mentioned impossibility of two-mode storage. %behavior described above.
%allowed to show that Stokes retrieval is determined primarily by signal input and that the strongest influence of the Stokes field on the signal occurs during the leakage.
Furthermore, we showed that a particularly convenient %the correct 
description of storage in an EIT-FWM system %must consider 
involves a joint signal-Stokes mode, whose dynamics we also studied.    %Additionally, we present a theoretical description of joint-mode dynamics.   
Quantum properties of our system %joint-mode 
will be investigated in the future using the analysis similar to that in Refs.\ \cite{harrisPRL06,kolchinPRA07,harrisPRL08}.  The authors thank M.\ D.\ Lukin for helpful discussions.  
%\textbf{cite Harris papers}.  
AVG wishes to acknowledge support from NSF grant PHY-0803371 and the Lee A. DuBridge Fellowship.  This research was supported by NSF grant PHY-0758010, Jeffress Research Grant J-847, and the College of William \& Mary.  

%%%%%%%%%%
%%%%%%%%%%
\appendix
\section{Derivation of Eqs.\ (\ref{Prapp1}), (\ref{Stapp1}), (\ref{Sapp1}), and (\ref{sapp})}
%%%%%%%%%%
%%%%%%%%%%

In the main text, we omitted the derivations of Eqs.\ \eqref{Prapp1}, \eqref{Stapp1}, \eqref{Sapp1}, and \eqref{sapp}.  In this Appendix, we present these derivations.

Since experiments and numerics show that turning the control field off and back on has a negligible effect on the fields except for a delay and spin wave decay during the storage time, we solve Eqs.\ (\ref{prEq}-\ref{Peq}) in the main text assuming a constant control field. In the co-moving frame ($\partial_t + c \partial_z \rightarrow c \partial_z$), Fourier transforming in time ($t \rightarrow \omega$ and $\partial_t \rightarrow - i \omega$), Eqs.\ (\ref{prEq}-\ref{Peq}) can be written as
\begin{widetext}
\begin{eqnarray}
\partial_z \left(\begin{array}{c}
\e(z,\omega) \\
\e'^*(z,\omega)
\end{array}
 \right) = i \frac{\alpha_0 \gamma}{2 F}\left( \begin{array}{cc}
\omega + i \Gamma_0 & - \frac{\Omega^2}{\Dhf} \\
\frac{\Omega^2}{\Dhf} & - \frac{\Omega^2}{\Dhf^2} (\omega + i \Gamma)
 \end{array}
 \right) \left(\begin{array}{c}
\e(z,\omega) \\
\e'^*(z,\omega)
\end{array}
 \right) = M \left(\begin{array}{c}
\e(z,\omega) \\
\e'^*(z,\omega)
\end{array}
 \right), \label{ap1}
\end{eqnarray}
\end{widetext}
where $F = \Omega^2 + (\Gamma - i \omega) (\Gamma_0 - i \omega)$ [see Eq.\ (2) in Ref.\ \cite{phillipsJMO09}].

To gain some intuition for how FWM may result in amplification, one can consider a simple case, in which the diagonal terms in the matrix $M$ in Eq.\ (\ref{ap1}) vanish (equivalently, one could consider the case where the Stokes field also propagates in its own EIT medium). Approximating further $F \rightarrow \Omega^2$, we find that
\begin{eqnarray}
M \approx  i \frac{\alpha_0 \gamma}{2 \Dhf}\left( \begin{array}{cc}
0 & -1 \\
1 & 0
 \end{array}
 \right)
\end{eqnarray}
and has eigenvectors $(1,\pm i)$ with eigenvalues $\pm \frac{\alpha_0 \gamma}{2 \Dhf}$, corresponding to an exponentially growing solution and an exponentially decaying solution. In our experiment, however, the diagonal terms for the signal and the Stokes fields are very different. Moreover, the effect of FWM is rather small and can, in fact, be treated perturbatively, as we will show below.

We checked numerically that the last entry in the matrix $M$ in Eq.\ (\ref{ap1}) does not significantly affect our results.  
For example, it gives a contribution to $\e'(L,\omega)$ of order $\alpha_0 L \gamma^2/\Dhf^2$, which will be negligible relative to other contributions of order $(\alpha_0 L)^2 \gamma^2/\Dhf^2$ since $\alpha_0 L \gg 1$. We will therefore neglect the last entry in the matrix $M$ in Eq.\ (\ref{ap1}) for the rest of this Appendix.

Eq.\ (\ref{ap1}) can then be solved  to give \cite{phillipsJMO09, lukinPRLFWM99, taoFWM, howellNatPh09,lukinPRL81}
\begin{widetext}
\begin{eqnarray}
 \left(\begin{array}{c}
\e(z,\omega) \\
\e'^*(z,\omega)
\end{array}
 \right) = e^{i \sigma z} \left( \begin{array}{cc}
\cosh(\xi z) + i \frac{\sigma}{\xi} \sinh(\xi z) & i \frac{2 \DR}{\beta} \sinh (\xi z) \\
-i \frac{2 \DR}{\beta} \sinh (\xi z) & \cosh(\xi z) - i \frac{\sigma}{\xi} \sinh(\xi z)
 \end{array}
 \right) \left(\begin{array}{c}
\e(0,\omega) \\
\e'^*(0,\omega)
\end{array}
 \right),\label{ap2}
\end{eqnarray}
%\end{widetext}
where $\DR = - \Omega^2/\Dhf$, $\beta = \sqrt{(\Gamma_0 - i \omega)^2 + 4 \DR^2}$, $\sigma = \frac{\alpha_0 \gamma}{4 F} (i \Gamma_0 + \omega)$, and $\xi = \frac{\alpha_0 \gamma}{4 F} \beta$.

Using the convolution theorem, we then obtain
%\begin{widetext}
\begin{eqnarray}
\e(z,t) &=& \int d t' \e(0, t - t') f_1(z,t') + \int d t' \e(0, t - t') f_2(z,t') + \int d t' \e'^*(0,t-t') f_3(z,t'),\label{eEqwithf}\\
 \e'^*(z,t) &=&  \e'^*(0, t) + \int d t' \e'^*(0, t - t')g_2(z,t') + \int d t' \e(0,t-t') g_3(z,t'), \label{ep}
\end{eqnarray}
%\end{widetext}
where
\begin{eqnarray}\label{f1expr}
f_1(z,t')  &=& \frac{1}{2 \pi} \int d \omega e^{2 i \sigma z} e^{- i \omega t'},\\
f_1(z,t') + f_2(z,t') &=& \frac{1}{2 \pi} \int d \omega e^{i \sigma z} \left[ \cosh(\xi z) + i \frac{\sigma}{\xi} \sinh(\xi z) \right]  e^{- i \omega t'},\\
f_3(z,t') &=& \frac{1}{2 \pi} \int d \omega e^{i \sigma z} i \frac{2 \DR}{\beta} \sinh(\xi z) e^{- i \omega t'},\\
\delta(t') + g_2(z,t') &=& \frac{1}{2 \pi} \int d \omega e^{i \sigma z} \left[ \cosh(\xi z) - i \frac{\sigma}{\xi} \sinh(\xi z)  \right] e^{- i \omega t'},\\
g_3(z,t') &=& - f_3(z,t').\label{g3expr}
\end{eqnarray}
\end{widetext}
%
%\begin{figure}[t]
%	\center{\includegraphics[width=\columnwidth]{figures/fandgfigs.eps}}
%	\caption{\label{fig:fandgfigs} Graphs of (a) $f_1(z=L,t)$, (b) $f_2(z=L,t)$, (c) $f_3(z=L,t)=-g_3(z=L,t)$, and (d) $g_2(z=L,t)$.  For sample parameters $\Omega=(2\pi) 10$ MHz, $\alpha_0 L=100$.  Red curves show the result of numerical integration of the respective expression in Eqs.~\eqref{f1expr}-\eqref{g3expr}.  Solid black curves show the approximate forms of the integrals, in Eqs. \eqref{f1}-\eqref{g3}.  The dashed black curves are the infinite-$\Ge$ approximations in Eqs. \eqref{f1}-\eqref{g3}}
%\end{figure}
%
\begin{figure*}[ht]
	\center{\includegraphics[width=\textwidth]{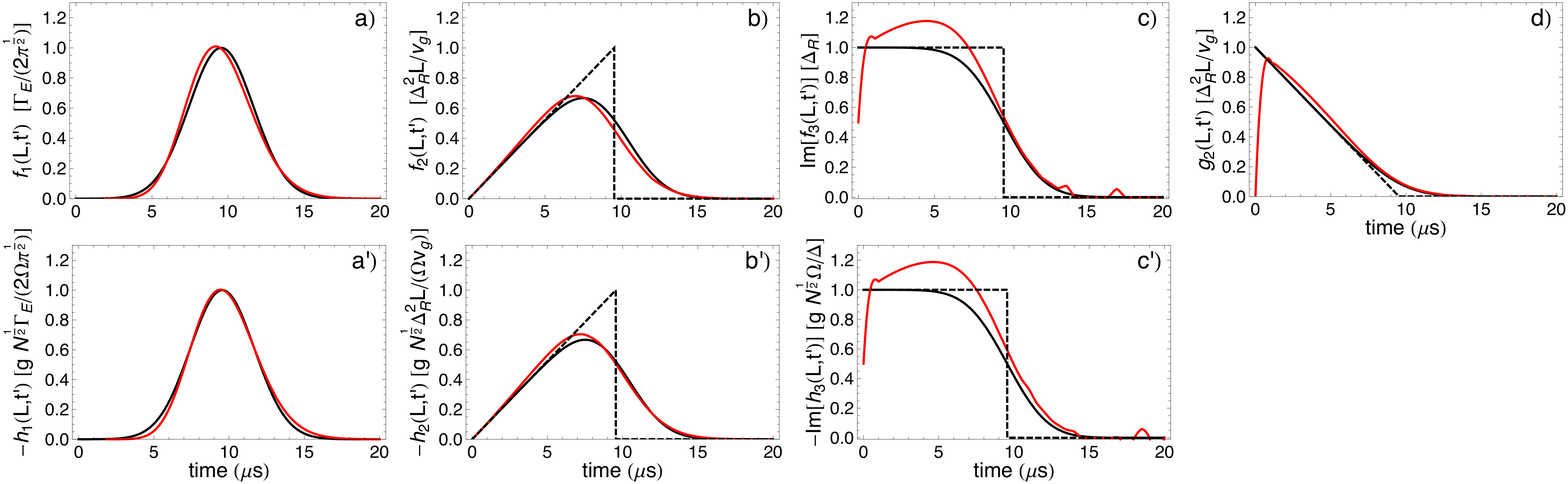}}
	\caption{\label{fig:fghfigs}(Color online) Graphs of (a) $f_1(L,t)$, (a$'$) $-h_1(L,t)$, (b) $f_2(L,t)$, (b$'$) $-h_2(L,t)$, (c) ${\rm Im}[f_3(L,t)]=-{\rm Im}[g_3(L,t)]$, (c$'$) $-{\rm Im}[h_3(L,t)]$, and (d) $g_2(L,t)$.   
For the calculations, $\Omega/(2\pi)=10$ MHz, $\alpha_0 L=80$.  Red curves show the result of numerical integration of the respective expression in Eqs.~(\ref{f1expr}--\ref{g3expr}, \ref{h1integral}--\ref{h3integral}).  Solid black curves show the approximate forms of the integrals given in Eqs.\ (\ref{f1}-\ref{g3}, \ref{happrox}), without taking the limit $\Ge \to \infty$.  The dashed black curves incorporate the $\Ge\to\infty$ approximations in Eqs.\ (\ref{f1}-\ref{g3}, \ref{happrox}).
}
%From MA file checkfandg.nb
\end{figure*}
Here $f_1$ and $f_2$ are defined in such a way that $f_1$ captures pure EIT, while $f_2$ describes how FWM changes the relationship between the input signal and the output signal. $f_3$ describes the effect of the input Stokes field on the output signal.  Similarly, the first term in Eq.\ (\ref{ep}) describes pure undistorted propagation of the Stokes field in the absence of FWM. $g_2$ describes how FWM changes the relationship between the input Stokes field and the output Stokes field. Finally, $g_3$ describes the effect of the input signal on the output Stokes field.

To get some insight into the behavior of $f_i$ and $g_i$, we consider the case $\delta = \delta_s$ (generalization to arbitrary $\delta$ is straightforward). We further take the limit $\gamma_0 = 0$, which is a reasonable approximation in our experiment, except during the waiting time between writing and retrieval (however, again one can easily generalize the derivation below to $\gamma_0 \neq 0$). Furthermore, we expand $f_2$ and $g_2$ to second order in $1/\Dhf$, and expand $f_3$ and $g_3$ to first order in $1/\Dhf$; in other words, we treat FWM perturbatively, which is a good approximation in our experiment, except in Figs.\ \ref{fig:stokesvsOD}(c, d). Furthermore, we approximate \cite{phillipsJMO09} $2 i \sigma \rightarrow i \frac{\omega}{v_g} - \frac{\omega^2}{L \Ge^2}$, where $v_g = \frac{2 \Omega^2}{\alpha_0 \gamma}$ is the EIT group velocity and $\Ge = \frac{\Omega^2}{\gamma \sqrt{\alpha_0 L/2}}$ is the width of the EIT transparency window.
%To get some insight, let's make approximations analogous to those in Eqs.\ (5,6) in JMO, but also keep the lowest-order nonzero terms in $f_2$ and $g_2$ (they will be of order $\DR^2$). In particular, in the off-diagonal term, we approximate $\beta \approx i \omega$ and $\xi \approx i \sigma$. In the diagonal terms, we approximate $\xi \approx i \sigma\left( 1 - \frac{2 \DR^2}{\omega^2}\right)$ and expand to first order in $\DR^2$. We can then approximate $2 i \sigma L \approx i \frac{\omega}{v_g/L} - \frac{\omega^2}{\Gamma_{E}^2}$. Then we take a lot of tricky integrals to obtain
We then find
\begin{widetext}
\begin{eqnarray}
f_1(z,t') &\approx& \frac{\Ge  e^{-\Ge^2 \frac{L}{4 z} (t'-z/v_g)^2}}{2 \sqrt{\pi z/L}} \approx \delta(t' - z/v_g),\label{f1}\\
f_2(z,t') &\approx& \DR^2 \left[ -\frac{ e^{-\Ge^2 \frac{L}{4 z} (t'-z/v_g)^2}}{2  \Ge \sqrt{\pi L/z}} + \frac{1}{2} |t'| +\frac{1}{2} t' \textrm{Erf}\left\{\frac{\Ge (z/v_g - t')}{2 \sqrt{z/L}}\right\}\right] \nonumber \\
& \approx& \DR^2 t' \Pi[0,z/v_g](t'),\\
f_3(z,t') &\approx& \frac{i \DR}{2} \left(\textrm{Sign}[t'] +  \textrm{Erf}\left\{\frac{\Ge  (z/v_g - t')}{2 \sqrt{z/L}}\right\}\right) \approx i \DR \Pi[0,z/v_g](t'),\label{f3}\\
g_2(z,t') &\approx& \DR^2  \left[- \frac{z  \delta(t') }{\Ge^2 L}+ \frac{ e^{-\Ge^2 \frac{L}{4 z} (t'-z/v_g)^2}}{\Ge \sqrt{\pi L/z} } + \frac{z/v_g-t'}{2}\left(\textrm{Erf}\left\{\frac{\Ge (z/v_g - t')}{2 \sqrt{z/L}}\right\} + \textrm{Sign}[t']\right)\right] \nonumber \\
&\approx & \DR^2 (z/v_g - t') \Pi[0,z/v_g](t'),\\
g_3(z,t') &= & - f_3(z,t').\label{g3}
\end{eqnarray}
\end{widetext}
Here Erf is the error function, the sign function Sign$[t'] = 1$ for $t' \geq 0$ and $-1$ otherwise, and the box function $\Pi[x,y](t) = 1$ for $x < t < y$ and $0$ otherwise.
%$f_1$ describes usual EIT. $f_2$, $f_3$, $g_2$, and $g_3$ describe FWM. In particular, $f_2$ describes the change in the signal from EIT to FWM with zero input Stokes. $f_3$ describes the effect of nonzero input Stokes on the signal. As expected, in the absence of FWM ($\DR = 0$), Stokes propagates undistorted at the speed of light. $g_2$ describes the change in the Stokes from free propagation to FWM with zero input signal. $g_3$ describes the effect of nonzero input signal on the Stokes.
The second approximation in Eqs.\ (\ref{f1}-\ref{g3}) is done in the limit $\Ge \rightarrow \infty$ (the case of an infinitely wide EIT window). Using the $\Ge \rightarrow \infty$ expressions, we arrive at Eqs.\ (\ref{Prapp1}, \ref{Stapp1}).  

In Fig.\ \ref{fig:fghfigs}(a-d), we plot functions $f_j$ and $g_j$ for $j=1,2,3$ and the two approximate forms described above.  Red curves depict the results of numerical integration of Eqs.\ (\ref{f1expr}-\ref{g3expr}), with experimental variables $\alpha_0 L=80$ and $\Omega/(2\pi)=10$ MHz, so that $\DR/(2\pi)=-14.6$ kHz, $\Ge/(2\pi)= 105$ kHz, and $v_g/(2\pi L)= 16.7$ kHz.  Because the light pulses $\e$ and $\e'^*$ have a finite bandwidth, we chose an integration bandwidth of $(2\pi) 160$ MHz, and we have checked that a larger range does not significantly affect the results.   Solid black curves in Fig.\ \ref{fig:fghfigs}(a-d) plot the first 
approximations in Eqs.\ (\ref{f1}--\ref{g3}); dashed black lines show the corresponding $\Ge\to\infty$ expressions in Eqs.\ (\ref{f1}--\ref{g3}).

Let us now compute $S(z,t)$. From Eqs.~\eqref{Seq} and \eqref{Peq}, we have
\begin{eqnarray}
S(z,\omega) = - \frac{g \sqrt{N} \Omega}{F} \left[\e(z,\omega) - i \frac{\Gamma-i\omega}{\Dhf} \e'^*(z,\omega)\right],
\end{eqnarray}
where $\e(z,\omega)$ and $\e'^*(z,\omega)$ are given in Eq.\ (\ref{ap2}).

We can then write
\begin{equation}
\begin{split}\label{Spinwithh}
S(z,t) =& \int d t' \e(0, t - t') h_1(z,t')\\
&+ \int d t' \e(0, t - t') h_2(z,t') \\
&+ \int d t' \e'^*(0,t-t') h_3(z,t'),
\end{split}
\end{equation}
where $h_1$ describes pure EIT, while $h_2$ and $h_3$ are the results of FWM.  Functions $h_j$ can be computed as
\begin{widetext}
\begin{eqnarray}\label{h1integral}
h_1(z,t') &=& \frac{1}{2 \pi} \int d \omega \frac{-g \sqrt{N} \Omega}{F} e^{2 i \sigma z} e^{- i \omega t'},\\
%h_2(z,t') & \approx & \frac{1}{2 \pi} \int d \omega \frac{g \sqrt{N} \Omega^3\left (F + e^{2 i \sigma z} (2 i \Omega^2 \sigma z - F)\right)}{F \Dhf^2 (\omega + i \Gamma_0)^2} e^{- i \omega t'},\\
%h_3(z,t') & \approx & \frac{1}{2 \pi} \int d \omega \frac{g \sqrt{N} \Omega \left(\Omega^2 e^{2 i \sigma z} - F\right)}{\Dhf (\omega + i \Gamma_0) F}  e^{- i \omega t'}.\label{h3integral}
h_1(z,t')+h_2(z,t')&=&\frac{1}{2 \pi} \int d \omega \frac{-g\sqrt{N} \Omega}{F} e^{i\sigma z}\left[\cosh(\xi z)+\left(i\frac{\sigma}{\xi}-\frac{2\DR(\Gamma-i\omega)}{\beta \Dhf}\right)\sinh(\xi z)\right]e^{- i \omega t'},\label{h2integral}\\
h_3(z,t')&=&\frac{1}{2 \pi} \int d \omega \frac{g\sqrt{N} \Omega}{F} e^{i\sigma z}\left[ i \frac{\Gamma-i\omega}{\Dhf}\cosh(\xi z)+\left(\frac{\sigma(\Gamma-i\omega)}{\xi \Dhf}-i\frac{2\DR}{\beta}\right)\sinh(\xi z)\right]\e^{- i \omega t'}.\label{h3integral}
\end{eqnarray}
%\end{widetext}

Expanding $h_2$ to $\mathcal{O}(1/\Dhf^2)$ and $h_3$ to $\mathcal{O}(1/\Dhf)$, the above expressions simplify to 
%
%\begin{widetext}
\begin{eqnarray}\label{h2h3approx}
h_2(z,t') & \approx & \frac{1}{2 \pi} \int d \omega \frac{g \sqrt{N} \Omega^3\left (F + e^{2 i \sigma z} (2 i \Omega^2 \sigma z - F)\right)}{F \Dhf^2 (\omega + i \Gamma_0)^2} e^{- i \omega t'},\\
h_3(z,t') & \approx & \frac{1}{2 \pi} \int d \omega \frac{g \sqrt{N} \Omega \left(\Omega^2 e^{2 i \sigma z} - F\right)}{\Dhf (\omega + i \Gamma_0) F}  e^{- i \omega t'}.
\end{eqnarray}
\end{widetext}

%Interestingly, it is immediately clear from Eq.\ (\ref{seq}) that, at this order in $1/\Delta$, FWM does not affect the contribution of $\e(0,t)$ to the spin wave.
Taking $\delta = \delta_s$, $\gamma_0 = 0$, $2 i \sigma \approx i \frac{\omega}{v_g} - \frac{\omega^2}{L \Gamma^2_E}$, and $F \approx \Omega^2$, %, and dropping $\omega (\omega + i \Gamma)$ in $h_2$,
we have
\begin{eqnarray}\label{happrox}
h_j(z,t') & \approx & - \frac{g \sqrt{N}}{\Omega} f_j(z,t')
%h_2(z,t') & \approx &  - \frac{g \sqrt{N}}{\Omega} f_2(z,t'),\\
%h_3(z,t') & \approx & - \frac{g \sqrt{N}}{\Omega} f_3(z,t'),
\end{eqnarray}
for $j = 1, 2, 3$, where the expressions for $f_j(z,t')$ are given in Eqs.\ (\ref{f1}-\ref{f3}).  Plugging Eq.\ \eqref{happrox} into Eq.\ (\ref{Spinwithh}) yields an expression that is proportional to the signal field $\e(z,t)$ in Eq.\ \eqref{eEqwithf}.  Thus, % to order $1/\Dhf^2$, 
under these approximations, 
we obtain Eq.\ (\ref{sapp}),
%\begin{equation}
%S(z,t) \approx - \frac{g \sqrt{N}}{\Omega} \e(z,t),\label{sapp}
%\end{equation}
which is, remarkably, the usual EIT relation.  Specifically, in the limit of an infinitely wide EIT window, $S(z,t)$ can be found by plugging Eq.\ (\ref{happrox}) [with the corresponding $\Ge \to \infty$ expressions for $f_j$ from Eqs.\ (\ref{f1}-\ref{f3})]  into Eq.\ (\ref{Spinwithh}) to yield Eq.\ \eqref{Sapp1}.   
The expressions for $|h_1(L,t)|$, $|h_2(L,t)|$, and $|{\rm Im}[h_3(L,t)]|$ are plotted in Fig.\ \ref{fig:fghfigs} (a$'$, b$'$, and c$'$), respectively.  As before, red traces depict the results of numerical integration of Eqs.\ (\ref{h1integral}--\ref{h3integral}) with the same input parameters as in the $f_j$ and $g_j$ analysis.  Solid black curves plot %the first approximation in 
Eq.\ \eqref{happrox} using finite $\Ge$ expressions for $f_j$ from Eqs.\ (\ref{f1}-\ref{f3}).  Dashed black lines show the corresponding $\Ge\to\infty$ expressions.


\begin{thebibliography}{00}
\bibitem{Omeganote}
As in Refs.\ \cite{fleischhauer,gorshkovPRL,gorshkovPRA2}, we define the Rabi frequency
$\Omega$ as $|\Omega|^2=\wp_{es}^2 I /(2 \hbar^2 \epsilon_0 c)$, where $I$ is
the control intensity.
\end{thebibliography}
\end{document}